\documentclass[superscriptaddress, aps, prl, twocolumn, showpacs]{revtex4}

\usepackage[pdftex]{graphicx}

\begin{document}

\title{Nodeless superconductivity arising from strong $(\pi,\pi)$ antiferromagnetism in the infinite-layer electron-doped cuprate Sr$_{1-x}$La$_{x}$CuO$_{2}$}

\author{John W. Harter}
\affiliation{Laboratory of Atomic and Solid State Physics, Department of Physics, Cornell University, Ithaca, New York 14853, USA}

\author{Luigi Maritato}
\affiliation{Department of Materials Science and Engineering, Cornell University, Ithaca, New York 14853, USA}
\affiliation{Universit\`{a} di Salerno and CNR-SPIN, 84084 Fisciano (SA), Italy}

\author{Daniel E. Shai}
\author{Eric J. Monkman}
\affiliation{Laboratory of Atomic and Solid State Physics, Department of Physics, Cornell University, Ithaca, New York 14853, USA}

\author{Yuefeng Nie}
\affiliation{Laboratory of Atomic and Solid State Physics, Department of Physics, Cornell University, Ithaca, New York 14853, USA}
\affiliation{Department of Materials Science and Engineering, Cornell University, Ithaca, New York 14853, USA}

\author{Darrell G. Schlom}  
\affiliation{Department of Materials Science and Engineering, Cornell University, Ithaca, New York 14853, USA}
\affiliation{Kavli Institute at Cornell for Nanoscale Science, Ithaca, New York 14853, USA}

\author{Kyle M. Shen}
\email[Author to whom correspondence should be addressed: ]{kmshen@cornell.edu}
\affiliation{Laboratory of Atomic and Solid State Physics, Department of Physics, Cornell University, Ithaca, New York 14853, USA}
\affiliation{Kavli Institute at Cornell for Nanoscale Science, Ithaca, New York 14853, USA}

\date{\today}

\begin{abstract}
The asymmetry between electron and hole doping remains one of the central issues in high-temperature cuprate superconductivity, but our understanding of the electron-doped cuprates has been hampered by apparent discrepancies between the only two known families: Re$_{2-x}$Ce$_{x}$CuO$_{4}$ and A$_{1-x}$La$_{x}$CuO$_{2}$. Here we report \textit{in situ} angle-resolved photoemission spectroscopy measurements of epitaxially-stabilized films of Sr$_{1-x}$La$_{x}$CuO$_{2}$ synthesized by oxide molecular-beam epitaxy. Our results reveal a strong coupling between electrons and $(\pi,\pi)$ antiferromagnetism that induces a Fermi surface reconstruction which pushes the nodal states below the Fermi level. This removes the hole pocket near $(\pi/2,\pi/2)$, realizing nodeless superconductivity without requiring a change in the symmetry of the order parameter and providing a universal understanding of all electron-doped cuprates.
\end{abstract}


\maketitle

In the hole-doped cuprates, it is well established that N\'{e}el antiferromagnetism (AF) is rapidly suppressed at a hole doping of $x \approx 0.03$, with $d$-wave superconductivity (SC) following at higher doping levels. In the electron-doped ($n$-type) materials, however, the situation is less clear \cite{armitage2010}. In the most studied $n$-type family, Re$_{2-x}$Ce$_{x}$CuO$_{4}$ (RCCO), where Re is a trivalent rare-earth cation, N\'{e}el AF persists up to an electron doping of $x = 0.14$ \cite{yamada2003, motoyama2007} before $d$-wave SC sets in. On the other hand, studies of the only other known $n$-type family, infinite-layer A$_{1-x}$La$_{x}$CuO$_{2}$ (A = Sr, Ca), suggest a less unified picture. A variety of probes including tunneling spectroscopy \cite{chen2002}, specific heat measurements \cite{liu2005}, neutron scattering \cite{white2008}, and muon spin rotation ($\mu$SR) \cite{khasanov2008} suggest a nodeless SC gap, in contrast to the hole-doped cuprates and RCCO. Moreover, zero-field $\mu$SR has not reported the presence of AF order in lightly doped  Sr$_{1-x}$La$_{x}$CuO$_{2}$ (SLCO) \cite{shengelaya2005}, and its SC has been shown to be dominated by electron-like rather than hole-like carriers \cite{fruchter2011}. 

In order to establish a unified understanding of the electron-doped cuprates, it is imperative to resolve the apparent discrepancies between these two families. Such efforts have been hampered by the inability to synthesize single crystals of A$_{1-x}$La$_{x}$CuO$_{2}$, precluding the most sophisticated probes of electronic and magnetic structure, such as angle-resolved photoemission spectroscopy (ARPES) and neutron diffraction. To overcome this barrier and resolve these long-standing questions, we have synthesized superconducting epitaxially-stabilized thin films of SLCO \cite{karimoto2001,leca2006,jovanovic2010} by oxide molecular-beam epitaxy (MBE) and probed their electronic structure \textit{in situ} with high-resolution ARPES. Our results demonstrate not only that robust $(\pi,\pi)$ AF is generic to the electron-doped cuprates, but that in SLCO the coupling of electrons to the AF is unusually strong, leading to a reconstruction of the Fermi surface (FS) in which the presumptive nodal hole pocket is pushed entirely below the Fermi energy ($E_{F}$). Removing the $d$-wave nodal states at $E_{F}$ from the normal state FS provides a natural mechanism for realizing nodeless SC without requiring a change in the symmetry of the order parameter away from $d$-wave.

\begin{figure*}
\includegraphics{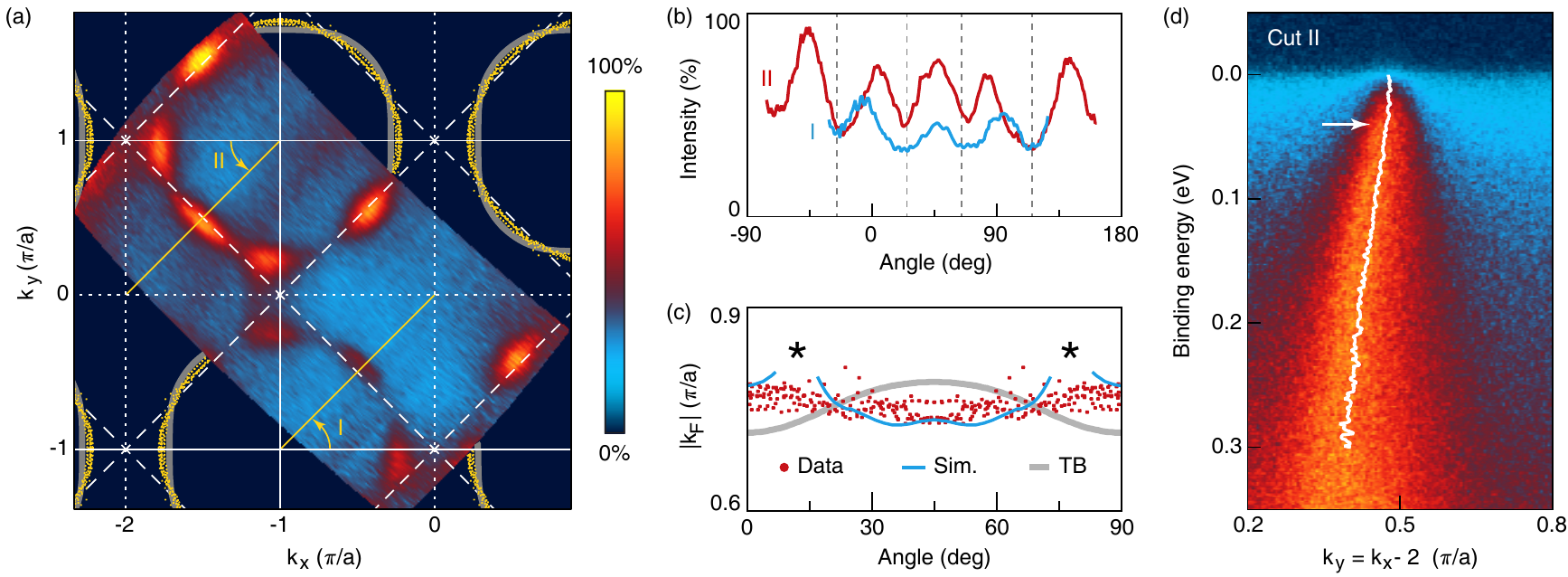}
\caption{\label{figure1}{(a) Unsymmetrized FS map for Sr$_{0.90}$La$_{0.10}$CuO$_{2}$ taken at 10 K showing spectral weight integrated within $E_{\mathrm{F}}$ $\pm$ 30 meV and normalized to a featureless background at high binding energy. Regions outside the map show the extracted $k_{\mathrm{F}}$'s (yellow points) and the TB FS (gray lines).  (b) Intensity versus angle along the two FS sheets, I and II, defined in panel (a). Dashed lines mark where the FS crosses the AF zone boundary.  (c) Experimental $k_{\mathrm{F}}$'s generated from the entire FS map and plotted by distance from $(\pi,\pi)$ as a function of angle around the FS. The points deviate significantly from the TB prediction but agree well with the simulation. The two asterisks mark the location of the hot spots, where a well-defined $k_{\mathrm{F}}$ cannot be reliably determined.  (d) Spectrum of nodal cut II in panel (a). The white line shows the band dispersion as determined by an MDC analysis. An upturn at 40 meV, also visible in Fig.\ \ref{figure2}(d), is identified by an arrow.}}
\end{figure*}

SLCO possesses a number of favorable characteristics that make it an ideal candidate for the study of cuprate SC. First, the crystal structure of SLCO is comprised solely of flat CuO$_2$ planes separated by alkaline earth atoms and is devoid of chains, orthorhombic distortions, incommensurate superstructures, ordered vacancies, and other complications that abound among the other cuprate families. In addition, SLCO has the highest $T_{c,\mathrm{max}}$ of all $n$-type cuprates, 43 K \cite{er1991}, and can also be hole-doped to a $T_{c,\mathrm{max}}$ of 110 K \cite{azuma1992}. Finally, it lacks the large rare-earth magnetic moments present in RCCO, which have been shown to couple to the magnetism in the CuO$_2$ plane \cite{lynn1990}. Epitaxial thin films of SLCO ($x = 0.10$) were grown on (110) GdScO$_{3}$ substrates using a Veeco GEN10 oxide MBE system. Shuttered layer-by-layer deposition was performed in distilled O$_{3}$ at a background pressure of $1 \times 10^{-6}$ Torr and monitored with reflection high-energy electron diffraction. Immediately after growth, the films were oxygen-reduced by vacuum annealing in the growth chamber and then transferred under ultra-high vacuum into the ARPES measurement chamber. Measurements were performed with a VG Scienta R4000 electron spectrometer and He-I$\alpha$ photons (21.2 eV) at a base temperature of 10 K. After ARPES, samples were characterized by x-ray diffraction, resistivity measurements using a Quantum Design Physical Properties Measurement System, x-ray absorption spectroscopy, and x-ray photoelectron spectroscopy.

In Fig.\ \ref{figure1}(a), we show a $\bf{k}$-resolved map of spectral weight near $E_{\mathrm{F}}$. A large circular Fermi surface centered at $(\pi,\pi)$, generic to all doped cuprates, is apparent. We extract Fermi wavevectors ($k_{\mathrm{F}}$'s) by fitting maxima in the momentum distribution curves (MDCs) used to generate the map. After applying appropriate symmetry operations, the set of $k_{\mathrm{F}}$'s are plotted as yellow points in the areas outside of the map. Also shown is a 2D tight-binding (TB) prediction for the FS. The parameters of the TB dispersion ($\mu$, $t$, $t'$, and $t''$) were generated by first constraining $\mu$ so that the electron filling was fixed at $x = 0.10$. Then $t'/t$ and $t''/t$ were varied to reproduce the shape of the FS predicted by band structure calculations (details can be found in the Supplementary Information). Finally, $t$ was adjusted to match the experimental high-energy dispersion in the nodal direction. The resulting TB parameters are $\mu = -6$ meV, $t = 215$ meV, $t' = -34$ meV, $t'' = 43$ meV.

Assuming a quasi-2D FS, the experimentally determined $k_{\mathrm{F}}$'s yield a Luttinger volume corresponding to $x = 0.09 \pm 0.02$, in agreement with the nominal doping level. Figures \ref{figure1}(b)-(d) show a number of notable features of the data. First, the intensity is strongly modulated as a function of angle around the FS, a phenomenon originally observed in Nd$_{1.85}$Ce$_{0.15}$CuO$_{4}$ \cite{armitage2001b}. Second, the location of spectral weight along the $(0,0)$--$(\pi,\pi)$ nodal direction deviates significantly from the TB prediction, whereas the agreement is better in other areas of momentum space. Third, the MDC-derived nodal quasiparticle (QP) dispersion shows a clear upturn at a binding energy of about 40 meV.

\begin{figure*}
\includegraphics{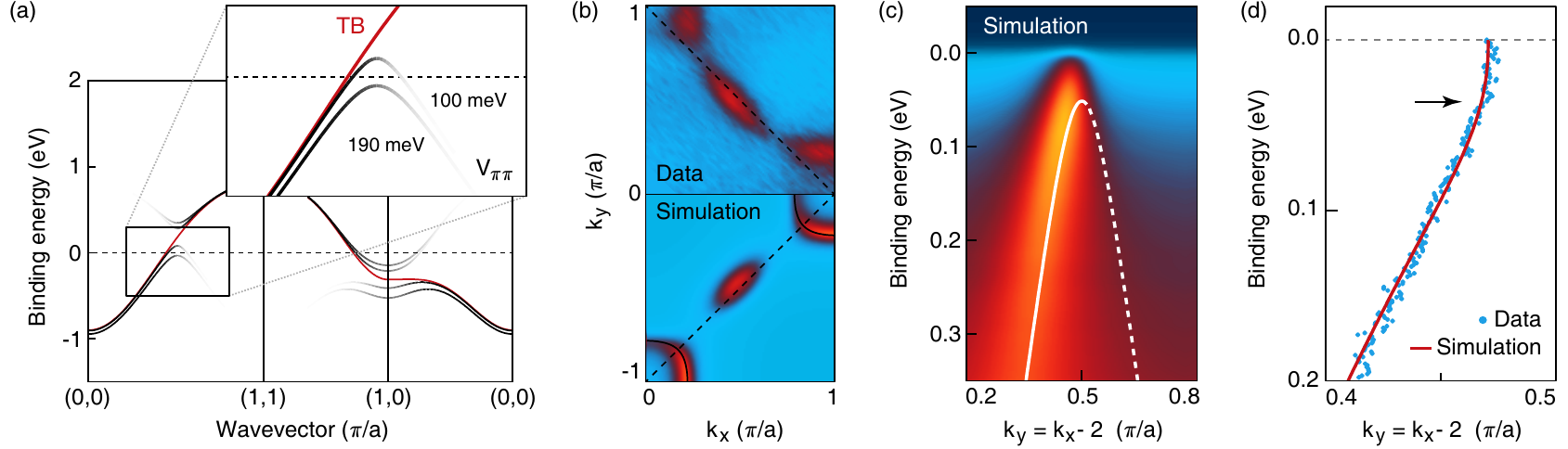}
\caption{\label{figure2}{(a) An illustration of the model discussed in the text. When $V_{\pi\pi} > 170$ meV, the nodal hole pocket is fully gapped.  (b) Comparison of the experimental FS map and the simulation, neglecting final-state photoemission matrix elements. The dashed line marks the AF zone boundary and the solid line shows the underlying FS of the model. In the data and simulation, the weight at $(\pi/2,\pi/2)$ is not due to a true band crossing, but instead comes from the tail of the broad QP spectral function.  (c) Simulation of the experimental spectrum presented in Fig.\ \ref{figure1}(d). The white line shows the model's underlying band structure.  (d) Nodal band dispersion, as determined by an MDC analysis. Dots are derived from the experimental data shown in Fig.\ \ref{figure1}(d) and the line is extracted from the simulation in panel (c) by an identical procedure.}}
\end{figure*}

To explain these features, we employ a simple model first, proposed for the RCCO family \cite{armitage2001b,matsui2005a,park2007}, whereby electrons with wavevectors \textbf{k} and \textbf{k} + $(\pi,\pi)$ are mixed via an off-diagonal matrix element $V_{\pi\pi}$. Despite the lack of explicit strong electron correlations, this model has been shown to be successful in reproducing the key low-energy features in the electronic structure of the RCCO family. The $V_{\pi\pi}$ term is assumed in this work to originate from static or slowly fluctuating AF. It could also arise from any sufficiently strong $(\pi,\pi)$ ordering \cite{chakravarty2001}, but the strong AF ordering in RCCO presents it as an obvious candidate. The term opens a gap of size $2\left|V_{\pi\pi}\right|$ at the intersection of the underlying band structure $\epsilon(\mathbf{k})$ with its image folded across the AF zone boundary $\epsilon'(\mathbf{k})$. The corresponding energies are:
$${E_{\pm}(\mathbf{k}) = \frac{\epsilon(\mathbf{k}) + \epsilon'(\mathbf{k})}{2} \pm \sqrt{\left(\frac{\epsilon(\mathbf{k}) - \epsilon'(\mathbf{k})}{2}\right)^2 + \left|V_{\pi\pi}\right|^2}}.$$
An illustration of this bandstructure is shown in Fig.\ \ref{figure2}(a). The gap results in so-called ``hot spots'' where spectral weight is dramatically suppressed, dividing the FS into two sheets: an electron pocket near the zone boundary at $(\pi,0)$, and a hole pocket in the nodal region at $(\pi/2,\pi/2)$. This readily explains the observed intensity modulation displayed in Fig.\ \ref{figure1}(b).

For $V_{\pi\pi}$ sufficiently large ($> 170$ meV in our model), the nodal pocket is pushed entirely below $E_{\mathrm{F}}$, leaving only an electron sheet around $(\pi,0)$. Such behavior has been reported for Sm$_{1.86}$Ce$_{0.14}$CuO$_{4}$ \cite{park2007} and Eu$_{1.85}$Ce$_{0.15}$CuO$_{4}$ \cite{ikeda2009}, and is consistent with our data. Figure \ref{figure4}(a) shows an AF-induced pseudogap in the energy distribution curve (EDC) at the node, and in Fig.\ \ref{figure2}(b)-(d), we compare simulations of the model for $V_{\pi\pi} = 190$ meV with our data (details can be found in the Supplementary Information), where we find that a fully gapped nodal pocket is consistent with all of the other features of the data. The submergence of the nodal pocket shifts the near-$E_{\mathrm{F}}$ intensity toward $(\pi/2,\pi/2)$, explaining the inconsistency highlighted in Fig.\ \ref{figure1}(c). Additionally, the upturn in the dispersion at 40 meV, as marked by the arrows in Fig.\ \ref{figure1}(d) and \ref{figure2}(d), is an artifact of the MDC analysis procedure in the presence of a gap and is commonly observed in other systems \cite{eschrig2002}; an identical MDC-analysis of our simulation yields a similar upturn. If we revise our earlier Luttinger count assuming only small electron pockets in a folded zone ($x$ vs. $1+x$), we obtain a doping $x = 0.10 \pm 0.03$, again consistent with our chemical composition. The ability to explain all experimental features using a simple model strongly suggests that the coupling of electrons to $(\pi,\pi)$ order in SLCO results in a reconstructed FS that gaps the nodal pocket. This cuprate FS topology, comprised solely of small electron pockets, has been reported in quantum oscillation measurements of YBa$_{2}$Cu$_{3}$O$_{6+\delta}$ at high magnetic fields \cite{doironleyraud2007} and may be relevant to those observations, particularly because such fields have been shown to stabilize AF \cite{lake2002,haug2009}.

\begin{figure}
\includegraphics{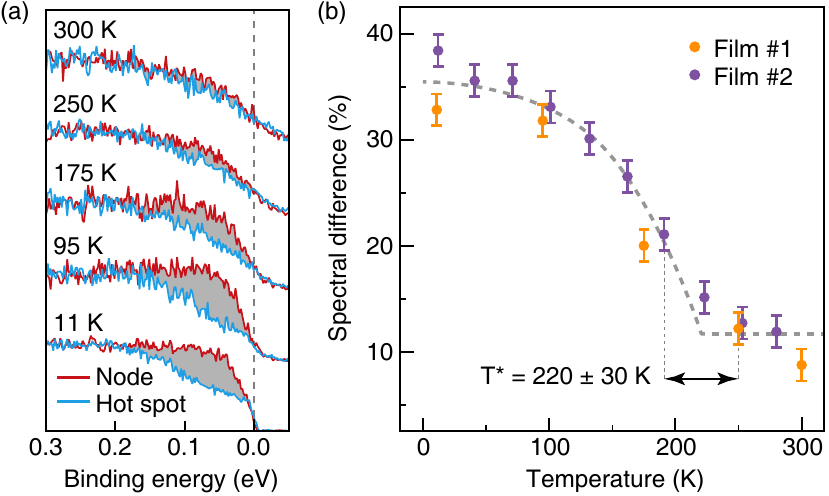}
\caption{\label{figure3}{(a) Temperature series of background-subtracted EDCs, offset for clarity, for Sr$_{0.90}$La$_{0.10}$CuO$_{2}$ at a node and a neighboring hot spot.  (b) Difference in spectral weight between the node and hot spot, shaded gray in panel (a), as a function of temperature.}}
\end{figure}

By comparing the near-$E_{\mathrm{F}}$ spectral difference between the node and the hot spot, we can remove trivial temperature effects from the Fermi step and determine the temperature dependence due to the AF. Figure \ref{figure3} shows this spectral difference as a function of temperature. Due to the presence of the AF gap, the intensity at the hot spot is dramatically reduced relative to the node over a 200 meV energy scale below $E_{\mathrm{F}}$, and this suppression drops rapidly between 150 K and 250 K. Assuming that static AF order exists, this could be associated with the closing of the gap above the N\'{e}el transition. From our data, we estimate a characteristic transition temperature $T^* = 220 \pm 30$ K. The similarity between our $T^*$ and the N\'{e}el temperature $T_{\mathrm{N}}$ for other electron-doped cuprates suggests that the observed spectral change could arise from the N\'{e}el transition. However, we cannot conclusively determine if the AF in SLCO is static or arises from fluctuating antecedent spin correlations.

\begin{figure}
\includegraphics{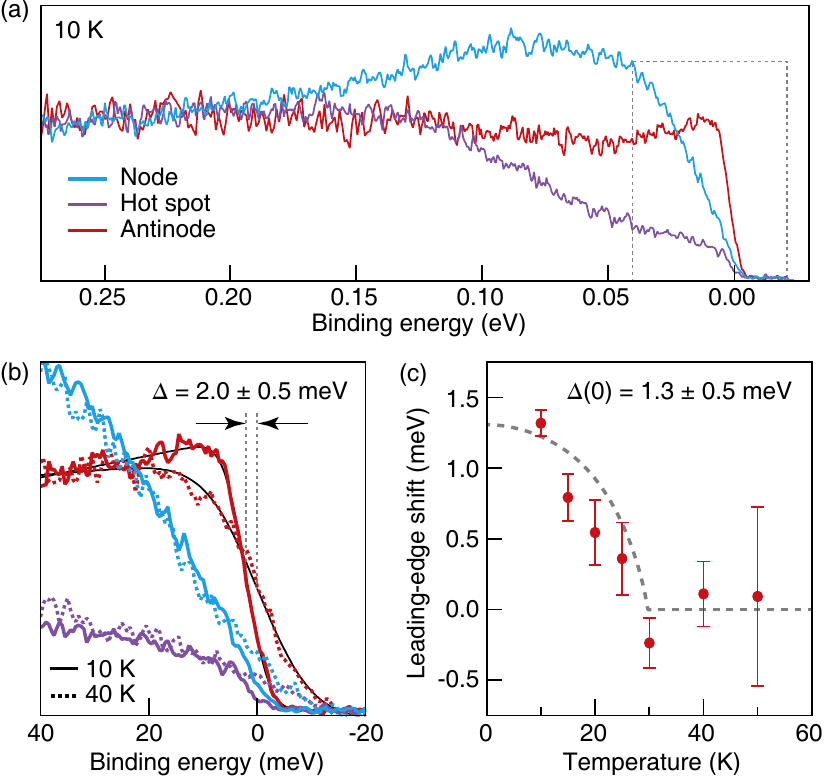}
\caption{\label{figure4}{(a) Comparison of low-temperature background-subtracted EDCs at the node, hot spot, and antinode for a superconducting Sr$_{0.90}$La$_{0.10}$CuO$_{2}$ film.  (b) Close-up of the dashed box in panel (a), with high-temperature data superimposed. The antinode shows a leading-edge shift at low temperature due to the SC gap. The solid black lines are fits to a Fermi function.  (c) Temperature dependence of the leading-edge shift for a second superconducting film.  An uncertainty of $\pm0.5$ meV for the SC gap $\Delta$ in both samples is estimated from systematic errors in background subtraction and the Fermi function fitting procedure.}}
\end{figure}

Measurements of non-superconducting (as-grown) Pr$_{2-x}$Ce$_{x}$CuO$_{4}$ have shown a gapping of nodal states argued to arise from the presence of excess oxygen \cite{richard2007}. Our samples were oxygen reduced and confirmed to be SC by \emph{ex situ} transport measurements with $T_c$'s in the range $25 \pm 5$ K. In addition, in Fig.\ \ref{figure4}, measurements from two samples exhibiting clear SC gaps of $\Delta$ = 1-2 meV on the electron pockets are shown, with the gap closing upon warming above $T_c$. This gap value is consistent with RCCO, where $\Delta \approx 2$ meV \cite{armitage2001a,diamant2009}, and confirms that small gaps are generic to electron-doped cuprates. As argued above, the hole pocket does not possess strong coherent weight at $E_{\mathrm{F}}$ and thus exhibits only a trivial temperature dependence. The gapping of the hole pocket by AF therefore can naturally explain the numerous reports of fully gapped SC in SLCO \cite{chen2002,liu2005,white2008,khasanov2008} without needing to invoke a change in the symmetry of the order parameter from $d$ to $s$. This nodeless $d$-wave scenario has been proposed theoretically by Yuan \textit{et al.} \cite{yuan2006} and Das \textit{et al.} \cite{das2007}, and coexisting AF and SC has been proposed theoretically by S\'{e}n\'{e}chal \textit{et al.} \cite{senechal2005}. Because the momentum range spanned by the electron pockets is narrow, we do not observe any substantial gap anisotropy, nor can we unequivocally rule out the possibility of $s$-wave SC. Our results demonstrate that cuprate high-$T_c$ SC can occur in a material with only electron-like carriers, coexistent AF, and without $d$-wave nodal QPs.

In conclusion, we have performed the first ARPES measurements on the infinite-layer cuprate Sr$_{1-x}$La$_{x}$CuO$_{2}$. Based on the accuracy of our Luttinger count, the success of our simple model, and the observation of a SC gap at $(\pi,0)$, we conclude that strong AF tendencies and SC coexist simultaneously and homogeneously in SLCO. Furthermore, the unusually strong coupling of electrons to $(\pi,\pi)$ AF results in a FS reconstruction comprised solely of electron-like carriers. SC is restricted only to electron pockets in SLCO, providing the first direct observation of high-$T_c$ SC in a cuprate completely devoid of hole-like carriers, as recently proposed by theoretical calculations \cite{yuan2006,das2007}. Furthermore, we have demonstrated that a gapping of the nodal states near $(\pi/2,\pi/2)$ by AF suppresses $d$-wave nodal QPs. This picture can provide a natural explanation of the earlier conflicting reports regarding the nature of SC in SLCO \cite{fruchter2011,chen2002,liu2005,white2008,khasanov2008,shengelaya2005}. By performing the first direct measurements of the electronic structure of an $n$-type cuprate distinct from Re$_{2-x}$Ce$_{x}$CuO$_{4}$, we have firmly established that robust AF and a small SC gap are intrinsic features of the electron-doped cuprates and not material-specific. 

We acknowledge helpful discussions with N.P.\ Armitage, J.C.S.\ Davis, and E.A.\ Kim, and thank B.\ Burganov and S.\ Chatterjee for assistance with measurements. This work was supported by the AFOSR (Grant No.\ FA9550-11-1-0033) and the National Science Foundation through the MRSEC program (Grant No.\ DMR-1120296). L.M.\ was supported by the ARO (Grant No.\ W911NF-09-1-0415) and E.J.M.\ acknowledges support from an NSERC PGS.

\end{document}